%% file: main.tex
\documentclass[twocolumn, 10pt, a4]{article}
\usepackage[a4paper, margin=1.8cm]{geometry}
\usepackage{graphicx} 
\usepackage{float}
\usepackage{subcaption}
\usepackage{cite}
\usepackage{amsmath} 
\usepackage{abstract}
\usepackage{amssymb} 
\usepackage{amsfonts} 
\usepackage{booktabs}
\usepackage{rotating}
\usepackage{hyperref}

\usepackage{etoolbox}
\usepackage[font={small}, labelfont={bf}]{caption}
\AtBeginEnvironment{tabular}{\small}

\usepackage[none]{hyphenat} 
\input{boxes.tex}
\title{A deep learning approach for marine snow synthesis and removal.}
\author{Fernando Galetto and Guang Deng}
\date{Nov 2023}

\begin{document}

\twocolumn[
\maketitle
  \begin{@twocolumnfalse}
\begin{abstract}
Marine snow, the floating particles in underwater images, severely degrades the visibility and performance of human and machine vision systems. This paper proposes a novel method to reduce the marine snow interference using deep learning techniques. We first synthesize realistic marine snow samples by training a Generative Adversarial Network (GAN) model and combine them with natural underwater images to create a paired dataset. We then train a U-Net model to perform marine snow removal as an image to image translation task. Our experiments show that the U-Net model can effectively remove both synthetic and natural marine snow with high accuracy, outperforming state-of-the-art methods such as the Median filter and its adaptive variant. We also demonstrate the robustness of our method by testing it on the MSRB dataset, which contains synthetic artifacts that our model has not seen during training. Our method is a practical and efficient solution for enhancing underwater images affected by marine snow.
\end{abstract}
\vspace{1cm}
  \end{@twocolumnfalse}
]

\section{Introduction}

The ability of machines to perceive the world as humans through imaging sensors has allowed researchers to create a massive number of tools to increase productivity, to improve performance and to solve important problems that couldn’t be solved in other way.  Factors such as noise, blurriness and low lighting conditions are the main enemies of computer vision algorithms and they are very common in underwater applications. 
Thus, underwater image enhancement is an important and challenging research topic that has been actively studied in recent years \cite{jaffe2014underwater, sheinin2016next}.

Underwater images and videos suffer from low visibility primarily due to scattering and absorption \cite{schettini2010underwater, ancuti2017color}. Absorption attenuates the light as it travels through  water, while scattering alters its direction \cite{ancuti2017color}. The presence of organic and inorganic matter in water contributes to both scattering and absorption, reducing visibility by attenuating light energy and deviating its trajectory. As depth and distance increase, wavelength-dependent attenuation leads to a particular color cast in underwater images \cite{mcglamery1980computer, blasinski2016three, akkaynak2017space, Akkaynak_2018_CVPR}.

Several methods have been developed to enhance and restore underwater images and videos, some of them require special hardware \cite{murez2015photometric, treibitz2012turbid, huang2016underwater, hu2018underwater, liu2019polarization} or multiple images \cite{roser2014simultaneous, wang2011research, zhang2014underwater} but single image methods are preferred due to their simplicity and adaptability to existing imaging systems.  Single image methods mostly tackle problems associated with colour cast and haze-like low contrast effect. Model-based methods use a physical model to describe the degradation and formulate the restoration as an inverse problem \cite{carlevaris2010initial,drews2013transmission,galdran2015automatic,peng2017underwater, akkaynak2019sea}. Some models used the traditional image formation model \cite{narasimhan2002vision} while others specifically developed for underwater scenarios \cite{mcglamery1980computer,blasinski2016three,akkaynak2017space, Akkaynak_2018_CVPR}. 

Machine learning methods were also proposed to enhance underwater images \cite{anwar2020diving}, the lack of ground truth images for underwater image enhancement made generative methods to stand up from the rest \cite{fabbri2018enhancingUGAN,guo2019underwaterDenseGAN, wang2019uwgan, lu2019multiCycleGAN, ye2018underwaterUIEsGAN, zhang2023underwater}. Alternative datasets were created trying to simulate the underwater environment \cite{sun2018deepP2P, uplavikar2019all,shin2016estimation, anwar2018deep,li2019underwater, wang2017deepUENET} enabling researchers to train other types of models such as encoder-decoder models \cite{ sun2018deepP2P, uplavikar2019all, gangisetty2022underwater}, CNN models \cite{shin2016estimation, anwar2018deep, li2019underwater} and other multi-branch architectures \cite{wang2017deepUENET} which performs very well under assumed conditions but are not robust for real-world applications. 

Floating particles, also known as marine snow, produce back-scattering causing a significant problem in real-life applications such as vessel hull cleaning and unmanned asset inspection. Despite recent advances in underwater image enhancement, only a limited number of proposed methods (that will be discussed in later sections) address it \cite{boffety2012phenomenological,boffety2012color, sato2021marine, banerjee2014elimination, farhadifard2017single, farhadifard2017marine, cyganek2018real, wang2021underwater, guo2022marine, jiang2020novel}. State-of-the-art results have not been achieved yet,  mainly due to the complexity of marine snow artifacts and the lack of realism produced by the existing models. To tackle this problem, we introduce a novel method based on a generative model to synthesize marine snow and a CNN model for image to image translation to reduce marine snow. Key contributions of this paper are as follows. 

\begin{itemize}

    \item GAN-based Marine Snow Synthesis: We present a Generative Adversarial Network (GAN) model capable of synthesizing samples of marine snow, replicating its complex characteristics.

    \item Paired Dataset Creation: We construct a dataset for marine snow removal by linearly combining natural underwater images with randomly distributed synthetic marine snow samples.

    \item CNN for marine snow removal: We propose a CNN architecture that effectively enhances underwater images by removing artifacts caused by marine snow. 

\end{itemize}

The paper is organized as follows. In Section 2, we provide an overview of the main characteristics of marine snow and review prior work related to marine snow removal. In Section 3, we present our method for synthesizing marine snow using the GAN model. The dataset and model for marine snow removal are described in Sections 4 and 5, respectively. Experimental results are presented in Section 6. Finally, we conclude our findings in Section 7.

\section{Previous work}

The back-scattering effect caused by floating particles, sediments, and bubbles is a widespread degradation problem that has been overlooked by most underwater image enhancement methods. This effect, significantly impacts image quality. Some studies have attempted to model marine snow by a simple Gaussian model \cite{boffety2012phenomenological,boffety2012color}. Sato et al. \cite{sato2021marine} categorized marine snow artifacts into two types and developed corresponding models to synthesize it. Unlike the Gaussian model, the proposed 3D plots resemble elliptic conical frusta, providing a fresh perspective on marine snow representation. 

For the removal of marine snow in images, methods based on median filter (MF) \cite{banerjee2014elimination, farhadifard2017single} have been used. However, the effectiveness of these methods is limited by the ability of the MF to remove large artifacts. In video processing, Farhadifard \cite{farhadifard2017marine} used background modeling to identify marine snow in static scenes, while Cyganek \cite{cyganek2018real} used a tracking method combined with MF. 

Recently, neural network-based methods have been studied. Koziarski et al. \cite{koziarski2019marine} trained a fully convolutional 3D neural network using manually labeled data to locate marine snow and combined it with an adaptive median filter to remove the artifacts. These video-based approaches may not be suitable for videos where numerous moving objects are present. The approach proposed in \cite{wang2021underwater} utilized three networks with the RESNET architectures and targeted fisheries videos. The method decomposed the input image into low and high-frequency components and applied separate networks for marine snow removal. However, its overall performance was not satisfactory. Guo et al.\cite{guo2022marine} treated the problem as an image-to-image translation problem. They created a dataset adding marine snow using Photoshop. Due to limited marine snow samples, the algorithm’s robustness was compromised. Jiang et al. proposed a different approach utilizing a GAN for denoising images affected by marine snow \cite{jiang2020novel}. The authors created a dataset by adding marine snow effect to underwater images from the IMAGENET dataset but the results on real underwater images were subpar, showing blurriness and incomplete artifact removal.

In summary, marine snow poses a significant challenge in underwater image processing. The existing approaches still require further improvement to achieve satisfactory results in real-life applications. A major limitation in the current studies is the scarcity of diverse marine snow samples, which hinders the performance of the proposed algorithms. Addressing this issue may lead to better marine snow removal techniques.

\begin{figure}[t]
    \centering
    \includegraphics[trim={0cm 0.3cm 0cm 2.5cm},clip, width=0.8\linewidth]{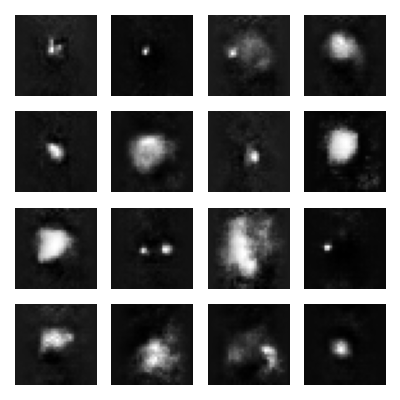}
    \caption{Synthetic marine snow samples.}
    \label{fig:marine_snow_samples}
\end{figure}

\section{ Synthesizing marine snow}

The appearance of marine snow can vary based on the scene's location and illumination, often leading to bright reflections when captured with a camera. Previous attempts at synthesizing marine snow using Gaussian functions or similar techniques lacked the realism required for training networks with robust performance. To overcome these limitations, we leverage the power of generative adversarial models, which have shown promise in learning and reproducing realistic samples. 

Our method begins with a dataset of natural underwater images, from which we extract and curate 2600 marine snow samples. Each sample is resized to a 32x32 patch size, and pixel values are scaled to a range from -1 to 1, optimizing suitability for training. Through this dataset, we train a generator model to produce fake samples of marine snow, while simultaneously training a discriminator model to distinguish between fake and real samples, resulting in an effective and visually convincing synthesis of marine snow. Figure \ref{fig:marine_snow_samples} shows 12 samples of marine snow produced by the generator after training for 10000 epochs.

\begin{figure}[t]
    \centering
    \includegraphics[width = 0.9\linewidth,keepaspectratio]{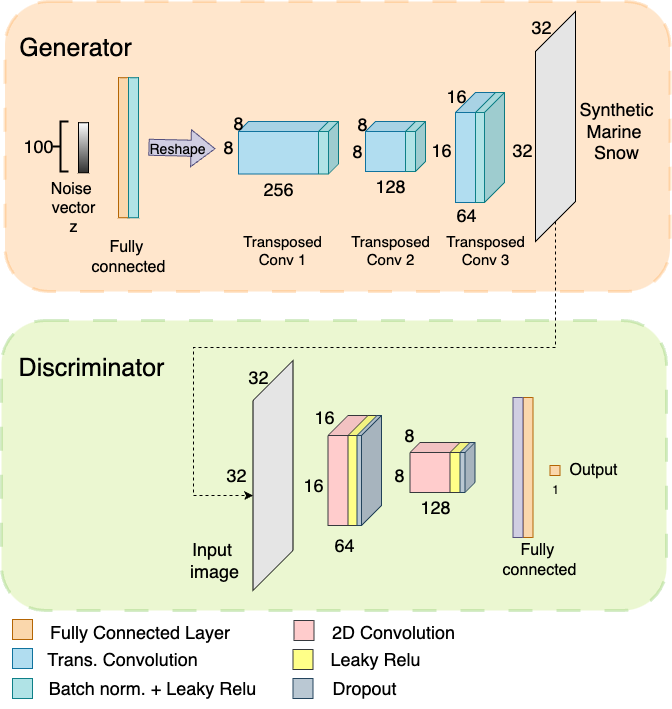}
    \caption{GAN architecture to synthesize marine snow artifacts. }
    \label{fig:GANarchitecture}
\end{figure}

The GAN architecture is shown in Figure \ref{fig:GANarchitecture}. The generator model is designed to produce realistic images based on a latent space representation. The model takes a 100-dimensional random noise vector $z$ as input and transforms it into a 32x32 grayscale image. It consists of several layers, including a dense layer, batch normalization, leaky ReLU activation, and convolutional transpose layers. The model progressively upscales the spatial dimensions of the tensor while reducing the number of channels. Batch normalization and leaky ReLU activation are applied after each transposed convolutional layer to improve training stability. The final convolutional transpose layer outputs a 32x32 image. The activation function used in the last layer is the hyperbolic tangent. Overall, this model demonstrates the ability to generate diverse and realistic images of marine snow from random noise. The discriminator model, which serves as the adversarial component, takes as input a 32x32 image and aims to distinguish between real and generated images. It consists of convolutional layers, each followed by leaky ReLU activation to introduce non-linearity. Dropout layers with a rate of 0.3 are added to prevent overfitting. The model further flattens the output and connects to a dense layer with a single output unit, responsible for making the decision on whether the input image is real or fake. The discriminator's role is to provide feedback to the generator to produce more realistic images.

The loss functions are part of the Wasserstein GAN (WGAN) formulation \cite{arjovsky2017wasserstein}, which provides better stability and convergence properties compared to the original objective function. The loss function for the discriminator (denoted as $\mathcal{L}_{\text{d}}$) and generator  (denoted as $\mathcal{L}_{\text{g}}$) are defined in Eq.\ref{eq:discrimnator_loss} and Eq.\ref{eq:generator_loss} respectively:

\begin{equation}
\mathcal{L}_{\text{d}} = \frac{1}{N} \sum_{i=1}^{N} \left( D(G(z_i)) - D(x_i) \right)    
\label{eq:discrimnator_loss}
\end{equation}
\begin{equation}
\mathcal{L}_{\text{g}} = -\frac{1}{N} \sum_{i=1}^{N} D(G(z_i))
\label{eq:generator_loss}
\end{equation}
where $D(G(z_i))$ is the output of the discriminator for the generated (fake) image $G(z_i)$, $D(x_i)$ is the output of the discriminator for the real image $x_i$, and $N$ is the batch size.

We followed the suggestion in \cite{arjovsky2017wasserstein} and used RMSprop\cite{tieleman2012lecture} as optimization method  with a learning rate of $5\times10^{-5}$ to stabilize the training process and mitigate some of the issues related to mode collapse and vanishing gradients.

\section{Dataset creation}\label{sec:Dataset}

\begin{figure}[t]
    \centering
    \includegraphics[width = 0.9\linewidth]{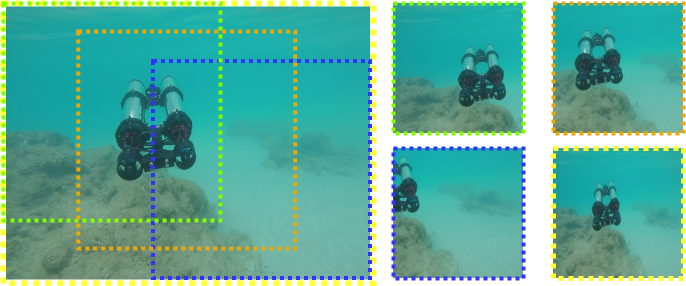}
    \caption{Cropping example}
    \label{fig:croppingimages}
\end{figure}

To train our marine snow removal model, we created a dataset using natural underwater images from three existing datasets: MSRB \cite{sato2021marine},  USR-248  \cite{islam2020underwater} and USOD \cite{islam2022svam}. The original images, used as ground truth, are free of marine snow. Because some images in the existing datasets are of high resolution, we derive three distinct images from each of these high-resolution images. We achieve this by cropping top-left, bottom-right, and center patches, all sized at our desired resolution of 384x384 pixels. Furthermore, we incorporate another image into this set by resizing the original image to match our target resolution. For a clear visual representation of this procedure, we refer to Figure \ref{fig:croppingimages}.

The dataset of images with marine snow is produced by linearly adding synthetic marine snow to the ground truth image. Specifically, let \(I \in R^{H\times W}\) be the ground truth and  \(P_i \in R^{m\times m}\) be a resized version of the $i$th synthetic marine snow sample. We add $N$ patches \(P_i\) to the ground truth at random positions \((x_i, y_i)\) resulting a distorted image $J$ :
\[ J = \min \left( 1 ,I + \sum_{i=1}^{N} \tau_i P_i \right),  \]
where $\tau_i$ is an attenuation coefficient, and the min-operation  enforces the condition $J\in[0,1]$.

\begin{figure}[t]
\centering

\subfloat[Original Image]{\begin{tikzpicture}[
            zoomboxarray,
            zoomboxarray columns=2,
            zoomboxarray rows=2,
            zoombox paths/.append style={line width=0.75pt}
        ]
            \node [image node] { \includegraphics[width=0.48\linewidth]{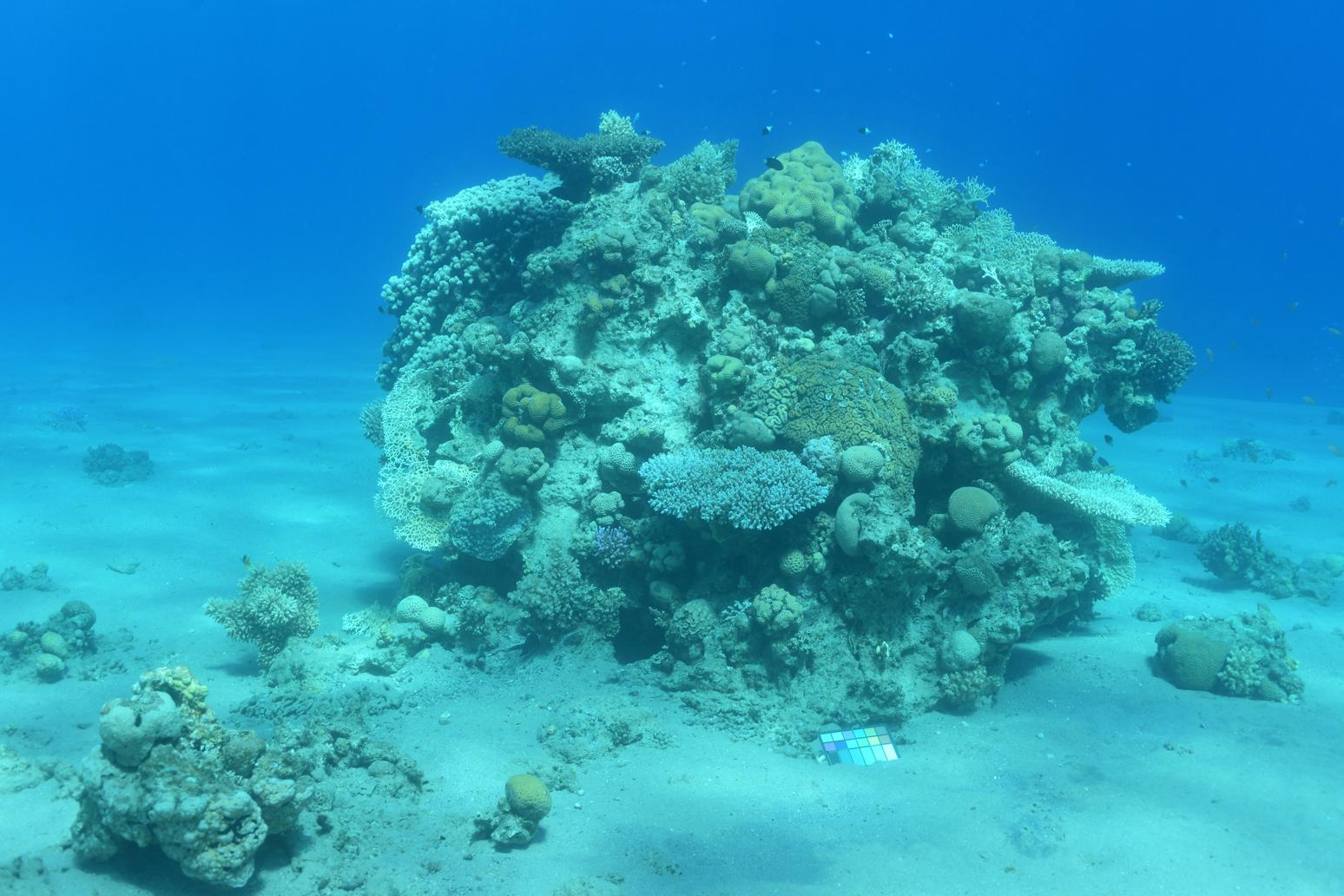} };

            \zoombox[color code = green,magnification=4]{0.1,0.6} 
            \zoombox[color code = orange,magnification=4]{0.9,0.8} 
            \zoombox[color code = red,magnification=4]{0.3,0.25}  
            \zoombox[color code = blue,magnification=4]{0.65,0.15} 
        \end{tikzpicture}}
        
\subfloat[Image with synthetic marine snow]{\begin{tikzpicture}[
            zoomboxarray,
            zoomboxarray columns=2,
            zoomboxarray rows=2,
            zoombox paths/.append style={line width=0.75pt}
        ]
            \node [image node] { \includegraphics[width=0.48\linewidth]{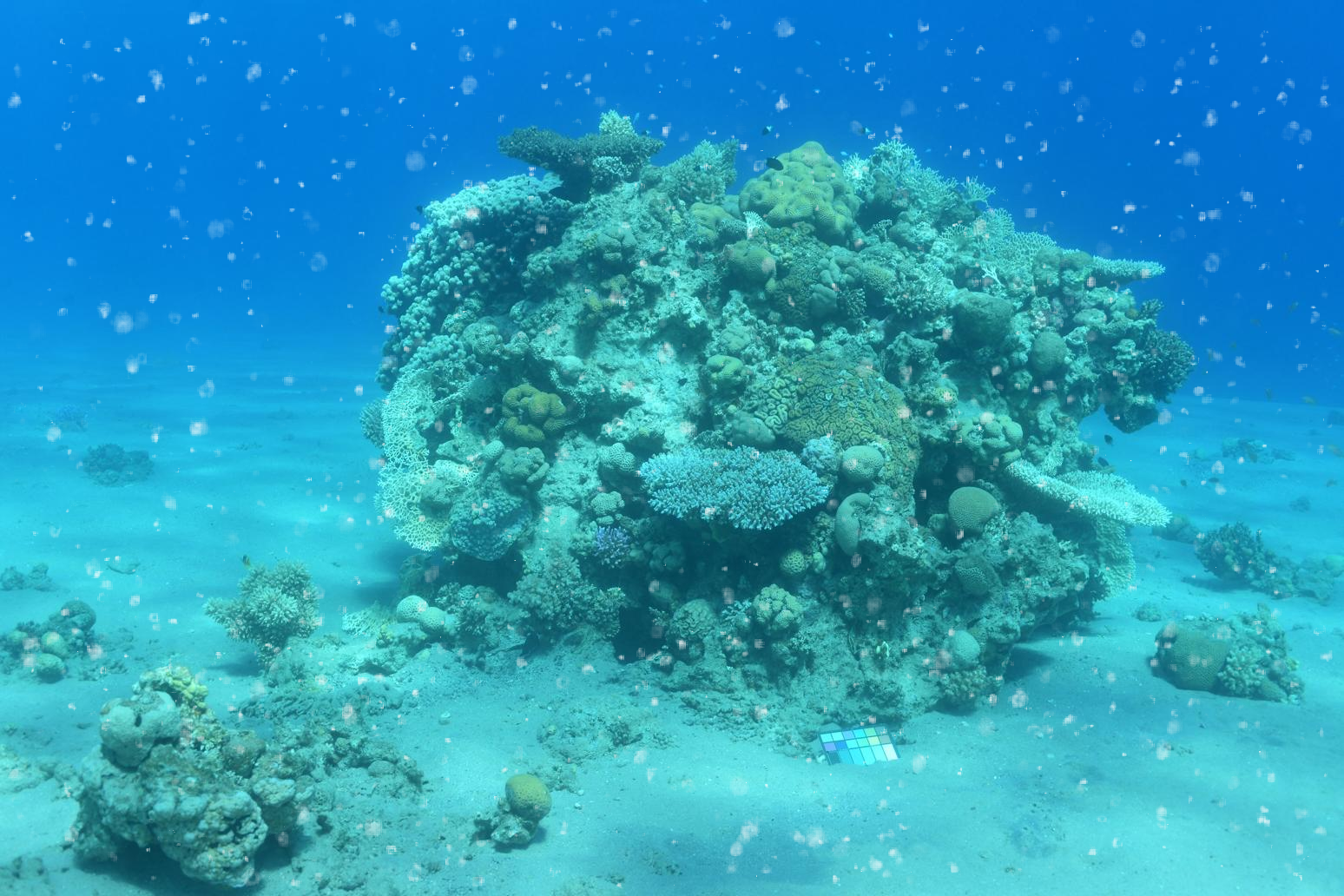} };

            \zoombox[color code = green,magnification=4]{0.1,0.6} 
            \zoombox[color code = orange,magnification=4]{0.9,0.8} 
            \zoombox[color code = red,magnification=4]{0.3,0.25}  
            \zoombox[color code = blue,magnification=4]{0.65,0.15} 
        \end{tikzpicture}}

\caption{Example of adding synthetic marine snow to a natural underwater image. }
\label{fig:distortedimagesample}
\end{figure}

In our experiment, we set the number of samples as $0 < N \leq 200$, the attenuation coefficient as $0.5 < \tau_i\leq 1.5$, and the patch size as $4 \leq m\leq 32$. These are random numbers from uniform distributions. Figure \ref{fig:distortedimagesample}a shows an example of a natural image without marine snow artifacts and Figure \ref{fig:distortedimagesample}b shows the result after placing 1000 samples of synthetic marine snow. 

After inspecting numerous images with marine snow, we observed that they usually have a significant amount of noise. So, to add realism to the generated image, we further added 3 different types of noise:

\begin{itemize}

    \item \textbf{Impulse noise:} It is used to model single pixel marine snow artifacts. This type of random noise manifests itself as isolated, randomly occurring bright pixels in the image.

    \item \textbf{Gaussian noise:} It is used to simulate random variations or errors in images. The Gaussian noise used in this paper has a standard deviation $\sigma^2= 10$ and mean $\mu=0$.

    \item \textbf{Poisson noise:} This type of noise is particularly prevalent in low-light conditions and is characterized by a single parameter denoted as $\lambda$. Experimentally, we found that $\lambda = 0.2$ produces realistic results.

\end{itemize}

Finally, we apply a data augmentation step by flipping each image horizontally, to increase the number of images in the dataset and avoid overfitting. We were able to create a dataset of 18846 paired color images that can be used to train and test a deep learning model for marine snow removal. We used 12869 images for training, 3217 for validation and 2760 for testing.

\section{Marine snow removal}

\begin{figure*}
    \centering
    \includegraphics[width =\linewidth]{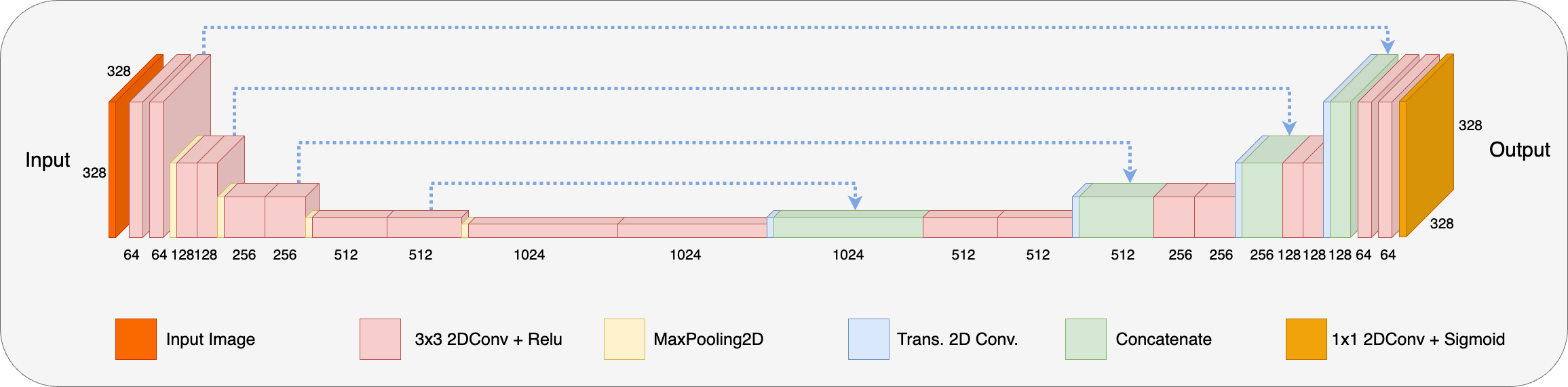}
    \caption{The Unet model architecture for marine snow removal.}
    \label{fig:UnetMarineSnow}
\end{figure*}

The dataset created in Section \ref{sec:Dataset} is used to train a deep learning model to remove the artifacts on images affected by marine snow. We employ a U-Net model architecture designed for image enhancement tasks. The U-Net architecture is shown in Figure \ref{fig:UnetMarineSnow}. The model follows an encoder-decoder structure with skip connections. The encoder path captures high-level features through multiple convolutional layers, max-pooling operations, and down sampling the spatial dimensions. This process helps the model learn significant image representations. The decoder path then uses transpose convolutions to up sample the feature maps and reconstruct the enhanced image with improved spatial details. The skip connections connect corresponding encoder and decoder layers, allowing the model to combine low-level and high-level features effectively. The final layer uses a 1x1 convolution with a sigmoid activation function to produce the enhanced image, preserving the color and spatial information. The U-Net architecture and its variants have demonstrated its effectiveness in improving the visual quality of images in various applications \cite{gangisetty2022underwater, sato2021marine}.

The model uses the mean squared error (MSE) and the perceptual loss functions. The perceptual loss leverages a pre-trained VGG19 neural network to extract high-level features from the true image $y$ and predicted enhanced image $\hat{y}$. By comparing these high-level features, the perceptual loss quantifies the perceptual similarity between the enhanced images and the ground truth. This approach aligns with the human visual perception, ensuring that the enhanced images preserve important visual characteristics and structural details.

The VGG19-based perceptual loss (denoted as $\mathcal{L}_{\text{p}}$) is calculated as the mean squared error (MSE) between the VGG feature maps:

\[
\mathcal{L}_{\text{p}}  = \frac{1}{N} \sum_{i=1}^{N} \left( VGG(y)_i - VGG(\hat{y})_i \right)^2
\]
where $N$ represents the total number of elements in the VGG feature maps, and $VGG(y)_i$ and $VGG(\hat{y})_i$ represent the $i$th element in the VGG feature maps of the ground truth and predicted images, respectively.

The perceptual loss encourages the U-Net model to generate predicted images that have similar high-level feature representations as the ground truth images, thereby capturing perceptual similarity between the two images rather than focusing solely on pixel-wise differences.

Additionally, the MSE loss (denoted as $\mathcal{L}_{\text{MSE}}$) is employed as a pixel-wise difference to capture the fine-grained differences between the true and the predicted enhanced images.  To compute the overall MSE loss for the entire image, we use the squared differences for all pixels:

\[
\mathcal{L}_{\text{MSE}}  = \frac{1}{N} \sum_{i=1}^{N}  \left( y_i - \hat{y}_i \right)^2
\]
where $N$ represents the total number of pixels in the image.

By combining both perceptual loss and MSE loss in the training process, the U-Net model is optimized to produce enhanced images that not only closely match the ground truth in terms of perceptual quality but also exhibit precise pixel-level similarities.  The combined loss is calculated as follows:

\begin{align}
   \mathcal{L}_{\text{U-Net}} &= \mathcal{L}_{\text{MSE}} + \gamma \mathcal{L}_{\text{p}} 
\end{align}
where $\gamma$ is a hyper-parameter that determines the relative importance of the perceptual loss compared to the MSE loss. By setting $\gamma=1$, we aim for a balanced trade-off between the pixel-wise accuracy and the preservation of high-level features in the generated images. Experimental results show that this setting leads to an effective and visually appealing marine snow removal.

\begin{figure}[h!]
    \centering
    \includegraphics[trim={1cm 0 1cm 0},clip, width = 0.8\linewidth]{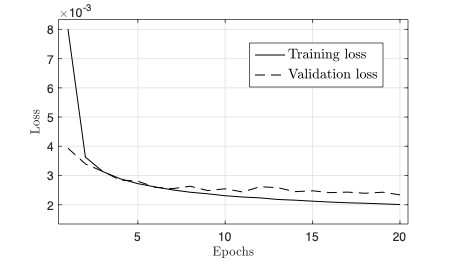}
    \caption{Training results for marine snow removal. Training and validation loss over epochs.}
    \label{fig:training-results}
\end{figure}

The model is trained using the Adam optimization algorithm configured with the following hyper-parameters: Learning rate (\(\alpha\)): 0.001, First moment decay rate (\(\beta_1\)): 0.9, Second moment decay rate (\(\beta_2\)): 0.999 and Epsilon (\(\epsilon\)): \(1 \times 10^{-7}\). Loss values per epochs are shown in Figure \ref{fig:training-results}. We can see that the training loss is 0.0020 and the validation loss is 0.0023 after 20 epochs. 

\section{Results}
In this section, we present the results of applying the trained U-Net to effectively remove marine snow from underwater images. We first evaluate the performance of our method by using the dataset described in Section \ref{sec:Dataset}, which encompasses images with synthetic marine snow. We then apply the U-Net to underwater images with real marine snow. Additionally, we demonstrated the utility of our method as a pre-processing step for enhancing underwater images (subsection 6.3). Finally, we evaluated our model using the benchmark proposed by Sato et al. \cite{sato2021marine}.

\subsection{Removing synthetic marine snow}

To assess the performance of our method, we compare it with the median filter, which effectively reduces impulsive noise while simultaneously preserving the sharpness of image edges. In this paper, we use kernel sizes of 3x3 and 5x5 pixels. Aiming for a more comprehensive comparison, we also include: BM3D (Block-Matching 3D) \cite{bm3d} and DnCNN (Denoising Convolutional Neural Network) \cite{zhang2017beyond} which are two different state-of-the-art image denoising techniques.  BM3D is a non-local image denoising algorithm that is particularly effective at removing noise from images while preserving important image structures and details. DnCNN is a deep learning-based image denoising technique that employs convolutional neural networks to learn the mapping from noisy images to clean images.

Results are summarized in Table \ref{tab:results}, which presents the average values of MSE, PSNR, and SSIM for each method. Remarkably, the proposed U-Net algorithm outperforms the Median filter across all metrics, indicating its superiority. The median filter excels in removing small artifacts with high intensities but underperforms in removing large artifacts. Larger kernel sizes could overcome this limitation at the cost of poor performance in edge preservation. The trained U-Net removes both small and large size artifacts while still preserving small details and sharp edges. 

BM3D and DnCNN, recognized for their efficacy in combating general noise types, prove less suitable for the unique challenges posed by marine snow. DnCNN, while effective in preserving small image features, fails in the removal of  medium and large artifacts. BM3D, shows some success in mitigating marine snow except in cases of high-intensity or larger artifacts but it sometimes eliminates fine details. These shortcomings make both BM3D and DnCNN less effective in the context of marine snow removal.

\begin{figure}[t!]
    \centering
     \hfill
     \subfloat[Distorted image]{\includegraphics[width=0.48\linewidth, height=0.35\linewidth, angle =180]{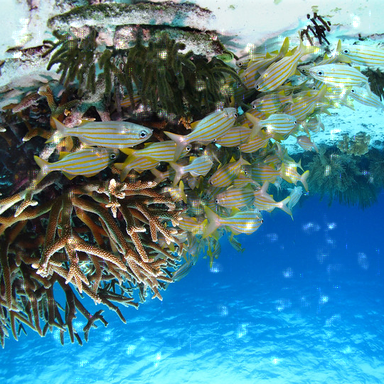}\label{subfig:img2}}
    \hfill
    \subfloat[Median Fil. 3x3]{\includegraphics[width=0.48\linewidth, height=0.35\linewidth, angle =180]{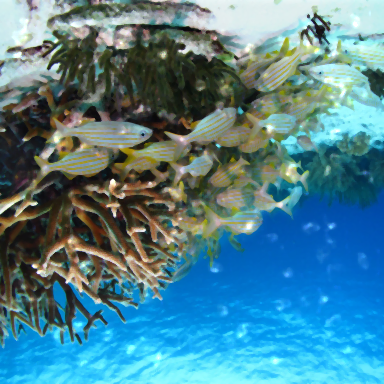}\label{subfig:img3}}
    \hfill

    \hfill
    \subfloat[Median Fil. 5x5]{\includegraphics[width=0.48\linewidth, height=0.35\linewidth, angle =180]{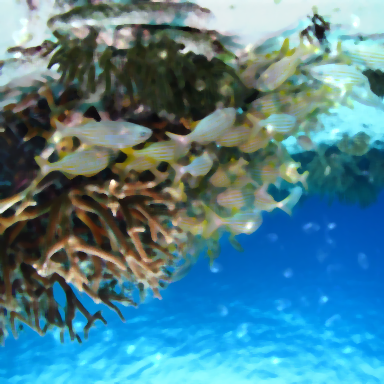}\label{subfig:img4bb}}
    \hfill
        \subfloat[DnCNN]{\includegraphics[width=0.48\linewidth, height=0.35\linewidth, angle =180]{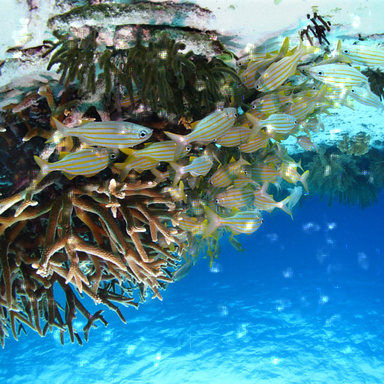}\label{subfig:img4vv}}
    \hfill

    \hfill
        \subfloat[BM3D]{\includegraphics[width=0.48\linewidth, height=0.35\linewidth, angle =180]{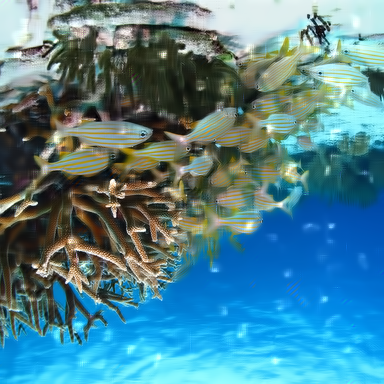}\label{subfig:imgdd4}}
    \hfill
    \subfloat[U-Net]{\includegraphics[width=0.48\linewidth, height=0.35\linewidth, angle =180]{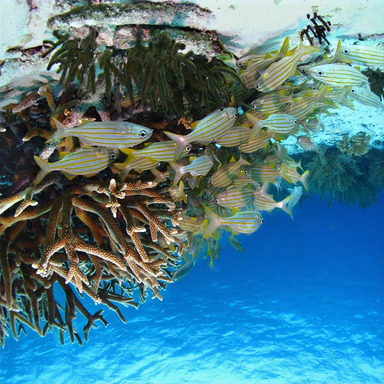}\label{subfig:img5}}
    \hfill
    \caption{Visual comparison example. Comparison results of removing synthetic marine snow with Median filter with kernel sizes of 3x3 and 5x5 and the proposed U-Net model for marine snow removal.  }
    \label{fig:results}

\end{figure}

\begin{table}[t]
\begin{center}
\caption{Results on the dataset with synthetic marine snow. Average MSE, PSNR and SSIM values for different methods.}
\label{tab:results}
\begin{tabular}{ l c c c }

\toprule
  & MSE & PSNR [dB] & SSIM \\
  \midrule
Test images & $59\times10^{-4}$ & 22.26 & 0.71\\  

 Med. Filt. $3\times 3$ & $47\times10^{-4}$ & 23.90 & 0.83\\  
 Med. Filt. $5\times 5$ & $56\times10^{-4}$ & 23.22 & 0.77\\  
 BM3D & $46\times10^{-4}$ &  23.95 &   0.88 \\
 DnCNN & $42\times10^{-4}$ &  24.50 &   0.91 \\
 U-Net & \boldmath{$13\times10^{-4}$} &  \textbf{29.18} &   \textbf{0.91} \\
 \bottomrule

\end{tabular}
\end{center}
\end{table}

To visualize the comparison effectively, Figure \ref{fig:results} demonstrates the visual output of both methods to remove synthetic marine snow. The graphical results substantiate the metrics, revealing that the proposed U-Net better preserves contrast and sharpness while successfully removing synthetic marine snow.

 \begin{figure*}[h!]
\centering
       
\subfloat[Distorted image]{\begin{tikzpicture}[
            zoomboxarray,
            zoomboxarray columns=2,
            zoomboxarray rows=1,
            zoombox paths/.append style={line width=0.75pt}
        ]
            \node [image node] {\includegraphics[trim={0 0 0 0},clip,width=0.24\linewidth,height=0.22\linewidth]{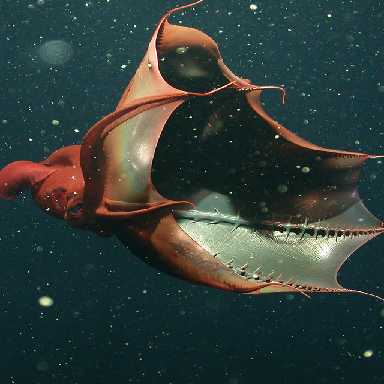} };

            \zoombox[color code = green,magnification=3]{0.3,0.8} 
            \zoombox[color code = orange,magnification=3]{0.85,0.5} 

        \end{tikzpicture}}
        \hfill
\subfloat[Median Fil. 3x3]{\begin{tikzpicture}[
            zoomboxarray,
            zoomboxarray columns=2,
            zoomboxarray rows=1,
            zoombox paths/.append style={line width=0.75pt}
        ]
            \node [image node] {\includegraphics[trim={0 0 0 0},clip,width=0.24\linewidth,height=0.22\linewidth]{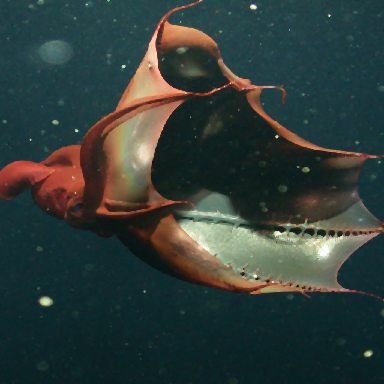}};

            \zoombox[color code = green,magnification=3]{0.3,0.8} 
            \zoombox[color code = orange,magnification=3]{0.85,0.5} 
        \end{tikzpicture}}

        \subfloat[Median Fil. 5x5]{\begin{tikzpicture}[
            zoomboxarray,
            zoomboxarray columns=2,
            zoomboxarray rows=1,
            zoombox paths/.append style={line width=0.75pt}
        ]
            \node [image node] {\includegraphics[trim={0 0 0 0},clip,width=0.24\linewidth,height=0.22\linewidth]{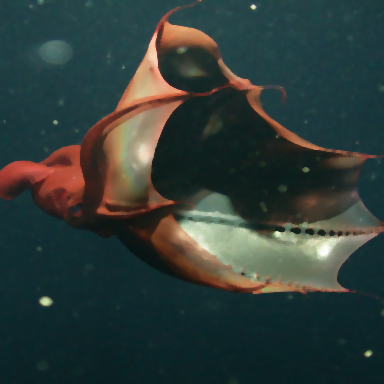}};

            \zoombox[color code = green,magnification=3]{0.3,0.8} 
            \zoombox[color code = orange,magnification=3]{0.85,0.5} 
        \end{tikzpicture}}
\hfill
            \subfloat[DnCNN]{\begin{tikzpicture}[
            zoomboxarray,
            zoomboxarray columns=2,
            zoomboxarray rows=1,
            zoombox paths/.append style={line width=0.75pt}
        ]
            \node [image node] {\includegraphics[trim={0 0 0 0},clip,width=0.24\linewidth,height=0.22\linewidth]{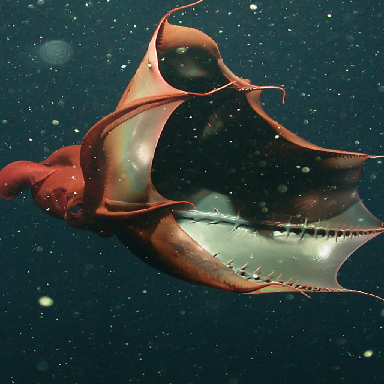} };

            \zoombox[color code = green,magnification=3]{0.3,0.8} 
            \zoombox[color code = orange,magnification=3]{0.85,0.5}  
        \end{tikzpicture}}
\hfill
        \subfloat[BM3D]{\begin{tikzpicture}[
            zoomboxarray,
            zoomboxarray columns=2,
            zoomboxarray rows=1,
            zoombox paths/.append style={line width=0.75pt}
        ]
            \node [image node] {\includegraphics[trim={0 0 0 0},clip,width=0.24\linewidth,height=0.22\linewidth]{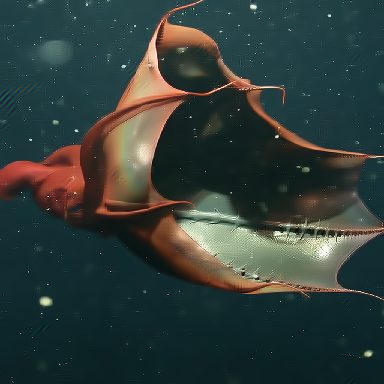} };

            \zoombox[color code = green,magnification=3]{0.3,0.8} 
            \zoombox[color code = orange,magnification=3]{0.85,0.5}  
        \end{tikzpicture}}
   \hfill   
           \subfloat[U-Net]{\begin{tikzpicture}[
            zoomboxarray,
            zoomboxarray columns=2,
            zoomboxarray rows=1,
            zoombox paths/.append style={line width=0.75pt}
        ]
            \node [image node] {\includegraphics[trim={0 0 0 0},clip,width=0.24\linewidth,height=0.22\linewidth]{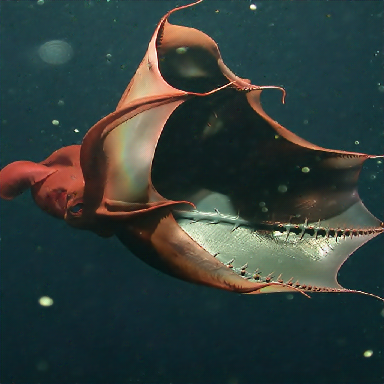} };

            \zoombox[color code = green,magnification=3]{0.3,0.8} 
            \zoombox[color code = orange,magnification=3]{0.85,0.5}  
        \end{tikzpicture}}

   \hfill   

    \caption{Visual comparison example. Comparison results of removing natural marine snow with Median filter with kernel sizes of 3x3 and 5x5 and the proposed U-Net model for marine snow removal.  }
\label{fig:resultsnat}
\end{figure*}
\begin{figure*}[t]
\hfill
     \subfloat[Input]{\includegraphics[width=0.32\linewidth, height=0.25\linewidth]{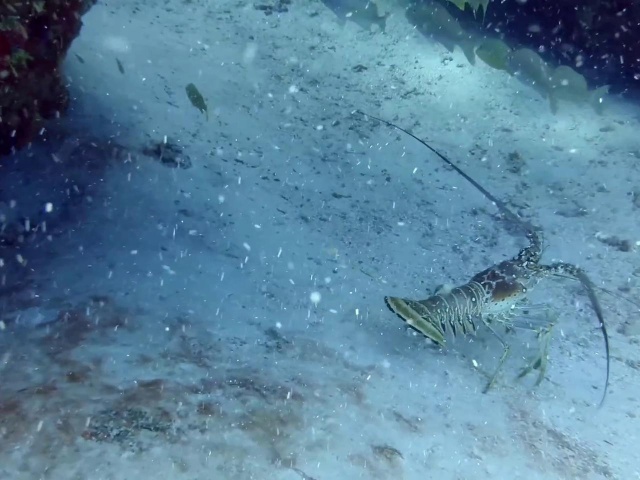}\label{subfig:o.mi}}
    \hfill
    \subfloat[Ancutti \cite{ancuti2017color}]{\includegraphics[width=0.32\linewidth, height=0.25\linewidth]{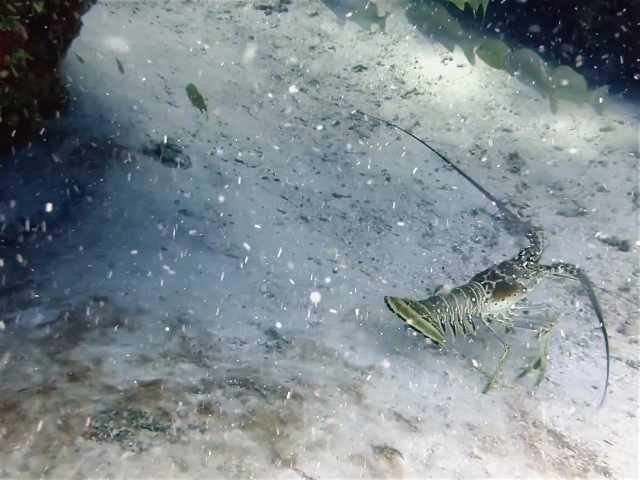}}
    \hfill
    \subfloat[Ancutti+U-Net]{\includegraphics[width=0.32\linewidth, height=0.25\linewidth]{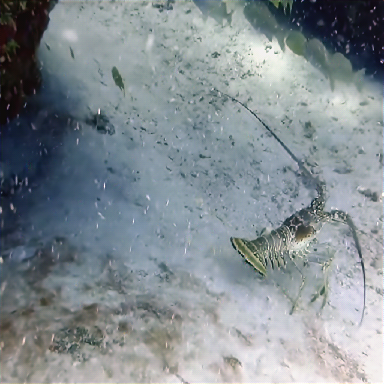}}
    \hfill
    \caption{Effect of removing marine snow on image enhancement.}
    \label{fig:resultsenhance}
\end{figure*}

\begin{table}[t]
\centering
\caption{Underwater image enhancement evaluation.}
\label{tab:metrics}
\begin{tabular}{lll}
\toprule
  & \textbf{UIQM} \cite{panetta2015human}$\uparrow$ & \textbf{UCIQE}\cite{yang2015underwater}$\uparrow$ \\
  \midrule
    \textbf{Original}  & 2.0677 & 0.5023 \\
    \textbf{Ancutti \cite{ancuti2017color}} &  3.3875& 0.5472 \\
    \textbf{Ancutti \cite{ancuti2017color}+Unet}  & \textbf{4.3477} & \textbf{0.5487 }\\
\bottomrule
\end{tabular}
\end{table}

\subsection{Removing natural marine snow}

Figure \ref{fig:resultsnat} illustrates a test aimed at showcasing the performance of the proposed U-Net in eliminating natural marine snow artifacts from real underwater images. The image was chosen because it contains a large amount of marine snow of different sizes and intensities. We can see that while the U-Net’s performance is not as good as that in removing synthetic marine snow, it still manages to significantly reduce the presence of natural marine snow. An important aspect to note is that the U-Net achieves this without sacrificing image details or textures through blurring. The visual contrast between the original and U-Net processed images is accentuated using colored rectangles (green and orange) to highlight the improvements.

We also compare results from the Unet with those obtained using a median filter of two distinct kernel sizes,  the BM3D, and DnCNN denoiser algorithms. The median filter, in both kernel sizes, effectively eliminates the presence of bright, high-intensity spots created by small and medium-sized objects while maintaining key edges and structures. However, it comes at the cost of losing fine details and textures, especially when using a larger kernel size. The U-Net closely matches the artifact-reduction capabilities of a 5x5 median filter but notably excels in preserving intricate image details, positioning it as a superior choice for marine snow removal.

DnCNN also preserves intricate image details but leaves most marine snow artifacts untouched. On the other hand, BM3D produces an evident reduction of artifacts in the image and effectively preserves details and edges. However, its performance is surpassed by the proposed U-Net especially when removing large size and bright artifacts. The U-Net removes a significant amount of marine snow artifacts from the original image, outperforming BM3D, retaining fine details and edges in the image.

\subsection{Underwater image enhancement}

There are a large number of methods for the enhancement and restoration of underwater images. Color cast and haze can be successfully removed. However, marine snow is not always considered by these methods, which not only fail to remove the artifacts but also fail to enhance the image when the marine snow is presented. In Figure \ref{fig:resultsenhance}, we demonstrate the performance of a state-of-the-art underwater image enhancement algorithm proposed by Ancutti et. all \cite{ancuti2017color}. As shown in Figure\ref{fig:resultsenhance}b, the enhancement algorithm reduces the color cast and improves the sharpness on the image. However, the sharpening is also applied to the marine snow artifacts, worsening its impact on the image quality. The marine snow artifacts have a higher intensity, making them more noticeable. 

To obtain the improved result shown in Figure\ref{fig:resultsenhance}c, we use the proposed U-Net to pre-process the input image before applying Ancutti’s algorithm. As can be seen, the algorithm still removes the color cast and improves the sharpness of the image, but the effect of marine snow is now notably reduced. 

A quantitative comparison is shown in Table \ref{tab:metrics}. We employ two well-known non-reference metrics widely employed to evaluate the quality of underwater images, UIQM \cite{panetta2015human} and UCIQE \cite{yang2015underwater}. UIQM is a comprehensive metric that measures sharpness, contrast and chromaticity to evaluate the quality of underwater images while UCIQE has been designed to emulate human quality perception. The metrics demonstrate that using the proposed U-Net as a pre-processing step with Ancuttis algorithms produces a higher quality image, improving the UIQM score by almost 30\% and preserving a similar UCIQE score.

\begin{figure*}[t!]
    \centering
{\includegraphics[width=0.195\linewidth]{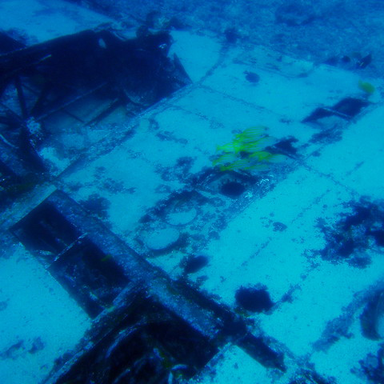}}\label{subfig:img1}
{\includegraphics[width=0.195\linewidth]{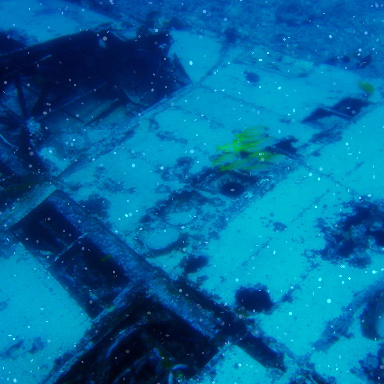}}\label{subfig:img2b}
{\includegraphics[width=0.195\linewidth]{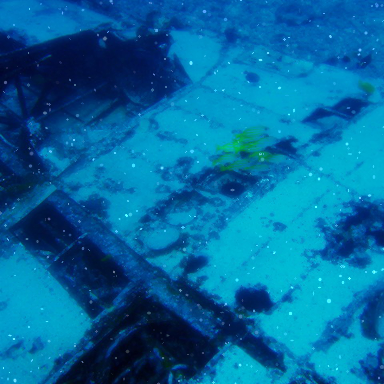}}\label{subfig:img3f}
{\includegraphics[width=0.195\linewidth]{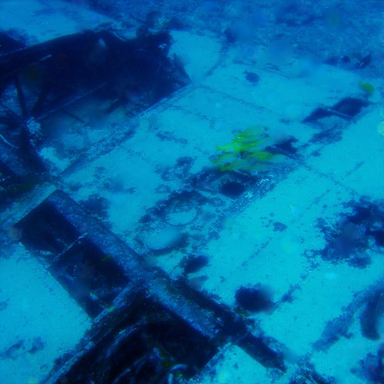}}\label{subfig:img4x}
{\includegraphics[width=0.195\linewidth]{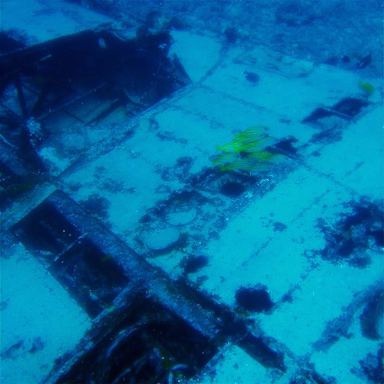}}\label{subfig:imgg5}
                
    \hfill
    \subfloat[Ground Truth]{\includegraphics[width=0.195\linewidth]{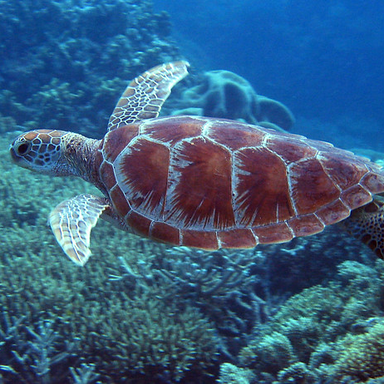}}\label{subfig:img6}
    \subfloat[Synthetic]{\includegraphics[width=0.195\linewidth]{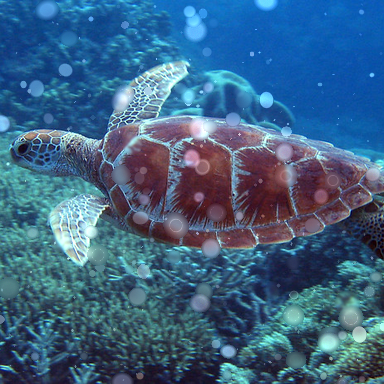}}\label{subfig:img7}
    \subfloat[Adapt. MF.]    {\includegraphics[width=0.195\linewidth]{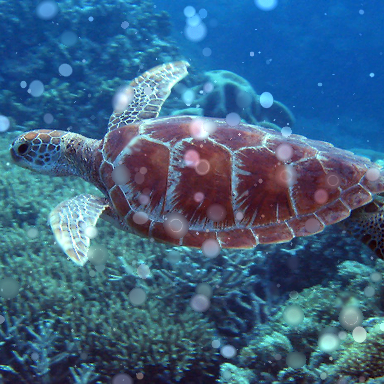}}\label{subfig:igmg81}
    \subfloat[MSRB]{\includegraphics[width=0.195\linewidth]{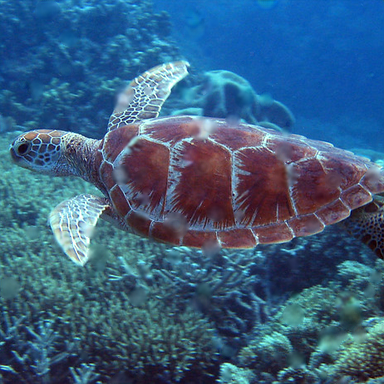}}\label{subfig:img8f1}
    \subfloat[Proposed]{\includegraphics[width=0.195\linewidth]{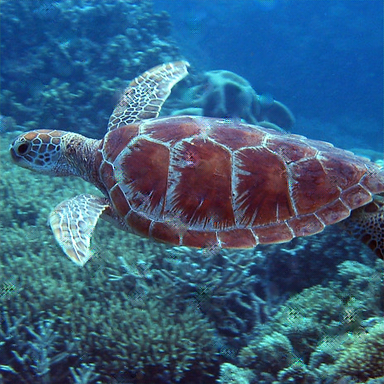}}\label{subfig:imdg81}
                
    \hfill

    \caption{Visual comparison example when removing synthetic marine snow from MSRB dataset. The First row shows as example of Task 1 from the benchmark proposed in \cite{sato2021marine}. Second row shows an example of Task 2.  }
    \label{fig:benchmark}
\end{figure*}

\begin{table}[h]
\begin{center}
\caption{Results on the MSRB test set. Average MSE, PSNR and SSIM values for different methods. }
\label{tab:resultsMSRB}
\begin{tabular}{ |l | c c | c c |}
\toprule
&  \multicolumn{2}{c|}{Task 1} & \multicolumn{2}{c|}{Task 2} \\
& PSNR   & SSIM  & PSNR & SSIM \\
\midrule
Test images  & 32.20 & 0.94 & 23.83 & 0.88\\  
Med.Filt.$3\times 3$ & 28.55 & 0.85 & 22.81 & 0.77\\  
Med.Filt.$5\times 5$  & 25.98 & 0.71& 21.93 & 0.64\\  
Adapt.Med.$3\times 3$ & 29.88 & 0.91 & 23.35 & 0.84\\  
Adapt.Med.$5\times 5$  & 28.08 & 0.86 & 22.83 & 0.79\\  
MSRB model  & 36.82 & 0.98 & 30.95 & 0.93\\  
Proposed &  \textbf{31.11} &   \textbf{0.94} & \textbf{27.27} & \textbf{0.90}\\
\bottomrule

\end{tabular}
\end{center}
\end{table}

\subsection{Comparison with MSRB}

We conduct a performance evaluation of the proposed U-Net using the benchmark framework introduced by Sato et al. in \cite{sato2021marine}. The primary objective of the benchmark is to assess the efficacy of the proposed U-Net in removing the presence of marine snow in each image from the MSRB (Marine Snow Removal Benchmark) dataset, followed by quantifying the quality of the denoised images using two essential metrics: PSNR and SSIM. For the assessment, undistorted images are utilized as reference.

The MSRB dataset incorporates synthetic marine snow. Diverging from conventional Gaussian models, the authors of this benchmark introduced a novel approach by representing the marine snow as 3D plots reminiscent of elliptic conical frustums. The benchmark comprises two distinct categories: Task 1 and Task 2. Task 1 involves images containing relatively smaller instances of synthetic marine snow, offering a challenging but manageable test. In contrast, Task 2 escalates the difficulty level, featuring images with marine snow samples of up to 32x32 pixels. 

A representative selection of images from the dataset, alongside their corresponding denoised outcomes, is presented in Figure \ref{fig:benchmark}. 
The top row shows an example image that belongs to the Task 1 test set, while the bottom row images belong to the Task 2 test set. As can be seen, our proposed U-Net successfully removes the synthetic marine snow produced by \cite{sato2021marine} in both Task 1 and Task 2. 

Table \ref{tab:resultsMSRB} presents the average PSNR and SSIM values obtained from our benchmark evaluation for both tasks. The proposed U-Net demonstrates strong performance when compared to the MSRB model proposed in \cite{sato2021marine}, and it outperforms both the median filter and its adaptive variants in this task.

It’s worth highlighting a crucial point: the MSRB model has been trained with synthetic data generated using the same methodology as the testing set. Hence, it is expected to excel in the task of removing such artifacts. On the other hand, our method performs well on this set, even though it was not trained on identical data. This showcases two significant findings. Firstly, our GAN model successfully generates diverse and realistic marine snow samples. Secondly, our proposed U-Net model demonstrates its capability to effectively identify and remove artifacts that are modeled as 3D plots reminiscent of elliptic conical frustums. These findings demonstrate the robustness and versatility of our approach.

We remark that it would be interesting to evaluate the performance of the MSRB model on other dataset such as the one that is created in this work. The result would then be used to compare the robustness of the MSRB model with the proposed U-Net mode.  However, we have not been able to run the MSRB model on our dataset since it is not publicly available.

\section{Conclusion}

In this paper, we proposed a novel approach to tackle the challenge of reducing the marine snow interference in underwater imagery. Our method involved the development of a WGAN model for the generation of realistic synthetic marine snow samples. These synthetic samples were then seamlessly integrated into real underwater images of diverse scene, forming a comprehensive dataset for marine snow removal. We trained a U-Net model on this dataset, using both Mean Squared Error (MSE) loss and perceptual loss. We showed that trained U-Net can remove synthetic marine snow to a high degree of accuracy. 

We conducted tests using the marine snow removal benchmark proposed by Sato et al. \cite{sato2021marine}. Despite not specifically training proposed U-Net model on their synthetic marine snow samples, results from the proposed U-Net demonstrated commendable performance, highlighting the potential of our GAN model in generating realistic synthetic marine snow compared to existing Gaussian models and the MSRB dataset.

A limitation of the proposed approach, which is in general associated with any data-driven approach, is that the performance of the resulting neural network is dependent on the training data to some extent. Further research could involve enriching WGAN model with a larger number of real marine snow samples to create an even more realistic dataset. This enhanced dataset, when used for retraining the U-Net, holds promise for improving the performance and adaptability to the nuances of marine snow removal in natural underwater environments.

\section*{Declarations}

\subsubsection*{Conflict of interest} 

The authors declare that they have no conflict of interest.

\subsubsection*{Data availability}

The datasets and models are publicly available in Github: \href{https://github.com/fergaletto/MSSR/}{http://github.com/fergaletto/MSSR/}.


\newpage

\input{moreresults}

\end{document}

%% file: boxes.tex
\usepackage{tikz}
\usepackage{pgfplots}
\pgfplotsset{compat=1.16}
\usetikzlibrary{spy,calc}

\newif\ifblackandwhitecycle
\gdef\patternnumber{0}
\pgfkeys{/tikz/.cd,
    zoombox paths/.style={
        draw=orange,
        very thick
    },
    black and white/.is choice,
    black and white/.default=static,
    black and white/static/.style={ 
        draw=white,   
        zoombox paths/.append style={
            draw=white,
            postaction={
                draw=black,
                loosely dashed
            }
        }
    },
    black and white/static/.code={
        \gdef\patternnumber{1}
    },
    black and white/cycle/.code={
        \blackandwhitecycletrue
        \gdef\patternnumber{1}
    },
    black and white pattern/.is choice,
    black and white pattern/0/.style={},
    black and white pattern/1/.style={    
            draw=white,
            postaction={
                draw=black,
                dash pattern=on 2pt off 2pt
            }
    },
    black and white pattern/2/.style={    
            draw=white,
            postaction={
                draw=black,
                dash pattern=on 4pt off 4pt
            }
    },
    black and white pattern/3/.style={    
            draw=white,
            postaction={
                draw=black,
                dash pattern=on 4pt off 4pt on 1pt off 4pt
            }
    },
    black and white pattern/4/.style={    
            draw=white,
            postaction={
                draw=black,
                dash pattern=on 4pt off 2pt on 2 pt off 2pt on 2 pt off 2pt
            }
    },
    zoomboxarray inner gap/.initial=5pt,
    zoomboxarray columns/.initial=2,
    zoomboxarray rows/.initial=2,
    subfigurename/.initial={},
    figurename/.initial={zoombox},
    zoomboxarray/.style={
        execute at begin picture={
            \begin{scope}[
                spy using outlines={%
                    zoombox paths,
                    width=\imagewidth / \pgfkeysvalueof{/tikz/zoomboxarray columns} - (\pgfkeysvalueof{/tikz/zoomboxarray columns} - 1) / \pgfkeysvalueof{/tikz/zoomboxarray columns} * \pgfkeysvalueof{/tikz/zoomboxarray inner gap} -\pgflinewidth,
                    height=\imageheight / \pgfkeysvalueof{/tikz/zoomboxarray rows} - (\pgfkeysvalueof{/tikz/zoomboxarray rows} - 1) / \pgfkeysvalueof{/tikz/zoomboxarray rows} * \pgfkeysvalueof{/tikz/zoomboxarray inner gap}-\pgflinewidth,
                    magnification=3,
                    every spy on node/.style={
                        zoombox paths
                    },
                    every spy in node/.style={
                        zoombox paths
                    }
                }
            ]
        },
        execute at end picture={
            \end{scope}
        },
        spymargin/.initial=0.5em,
        zoomboxes xshift/.initial=1,
        zoomboxes right/.code=\pgfkeys{/tikz/zoomboxes xshift=1},
        zoomboxes left/.code=\pgfkeys{/tikz/zoomboxes xshift=-1},
        zoomboxes yshift/.initial=0,
        zoomboxes above/.code={
            \pgfkeys{/tikz/zoomboxes yshift=1},
            \pgfkeys{/tikz/zoomboxes xshift=0}
        },
        zoomboxes below/.code={
            \pgfkeys{/tikz/zoomboxes yshift=-1},
            \pgfkeys{/tikz/zoomboxes xshift=0}
        },
        caption margin/.initial=0.01ex,
    },
    adjust caption spacing/.code={},
    image container/.style={
        inner sep=0pt,
        at=(image.north),
        anchor=north,
        adjust caption spacing
    },
    zoomboxes container/.style={
        inner sep=0pt,
        at=(image.north),
        anchor=north,
        name=zoomboxes container,
        xshift=\pgfkeysvalueof{/tikz/zoomboxes xshift}*(\imagewidth+\pgfkeysvalueof{/tikz/spymargin}),
        yshift=\pgfkeysvalueof{/tikz/zoomboxes yshift}*(\imageheight+\pgfkeysvalueof{/tikz/spymargin}+\pgfkeysvalueof{/tikz/caption margin}),
        adjust caption spacing
    },
    calculate dimensions/.code={
        \pgfpointdiff{\pgfpointanchor{image}{south west} }{\pgfpointanchor{image}{north east} }
        \pgfgetlastxy{\imagewidth}{\imageheight}
        \global\let\imagewidth=\imagewidth
        \global\let\imageheight=\imageheight
        \gdef\columncount{1}
        \gdef\rowcount{1}
        
    },
    image node/.style={
        inner sep=0pt,
        name=image,
        anchor=south west,
        append after command={
            [calculate dimensions]
            node [image container,subfigurename=\pgfkeysvalueof{/tikz/figurename}-image] {\phantomimage}
            node [zoomboxes container,subfigurename=\pgfkeysvalueof{/tikz/figurename}-zoom] {\phantomimage}
        }
    },
    color code/.style={
        zoombox paths/.append style={draw=#1}
    },
    connect zoomboxes/.style={
    spy connection path={\draw[draw=none,zoombox paths] (tikzspyonnode) -- (tikzspyinnode);}
    },
    help grid code/.code={
        \begin{scope}[
                x={(image.south east)},
                y={(image.north west)},
                font=\footnotesize,
                help lines,
                overlay
            ]
            \foreach \x in {0,1,...,9} { 
                \draw(\x/10,0) -- (\x/10,1);
                \node [anchor=north] at (\x/10,0) {0.\x};
            }
            \foreach \y in {0,1,...,9} {
                \draw(0,\y/10) -- (1,\y/10);                        \node [anchor=east] at (0,\y/10) {0.\y};
            }
        \end{scope}    
    },
    help grid/.style={
        append after command={
            [help grid code]
        }
    },
}

\newcommand\phantomimage{%
    \phantom{%
        \rule{\imagewidth}{\imageheight}%
    }%
}
\newcommand\zoombox[2][]{
    \begin{scope}[zoombox paths]
        \pgfmathsetmacro\xpos{
            (\columncount-1)*(\imagewidth / \pgfkeysvalueof{/tikz/zoomboxarray columns} + \pgfkeysvalueof{/tikz/zoomboxarray inner gap} / \pgfkeysvalueof{/tikz/zoomboxarray columns} ) + \pgflinewidth
        }
        \pgfmathsetmacro\ypos{
            (\rowcount-1)*( \imageheight / \pgfkeysvalueof{/tikz/zoomboxarray rows} + \pgfkeysvalueof{/tikz/zoomboxarray inner gap} / \pgfkeysvalueof{/tikz/zoomboxarray rows} ) + 0.5*\pgflinewidth
        }
        \edef\dospy{\noexpand\spy [
            #1,
            zoombox paths/.append style={
                black and white pattern=\patternnumber
            },
            every spy on node/.append style={#1},
            x=\imagewidth,
            y=\imageheight
        ] on (#2) in node [anchor=north west] at ($(zoomboxes container.north west)+(\xpos pt,-\ypos pt)$);}
        \dospy
        \pgfmathtruncatemacro\pgfmathresult{ifthenelse(\columncount==\pgfkeysvalueof{/tikz/zoomboxarray columns},\rowcount+1,\rowcount)}
        \global\let\rowcount=\pgfmathresult
        \pgfmathtruncatemacro\pgfmathresult{ifthenelse(\columncount==\pgfkeysvalueof{/tikz/zoomboxarray columns},1,\columncount+1)}
        \global\let\columncount=\pgfmathresult
        \ifblackandwhitecycle
            \pgfmathtruncatemacro{\newpatternnumber}{\patternnumber+1}
            \global\edef\patternnumber{\newpatternnumber}
        \fi
    \end{scope}
}

%% file: moreresults.tex
\onecolumn
\section*{Additional Results}
\begin{figure*}[h!]
     \subfloat{\includegraphics[width=0.24\linewidth, height=0.24\linewidth]{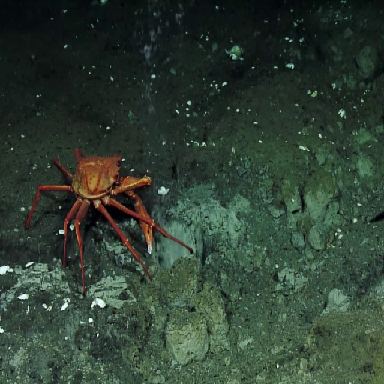}\label{subfig:ommi}}
    \hfill
    \subfloat{\includegraphics[width=0.24\linewidth, height=0.24\linewidth]{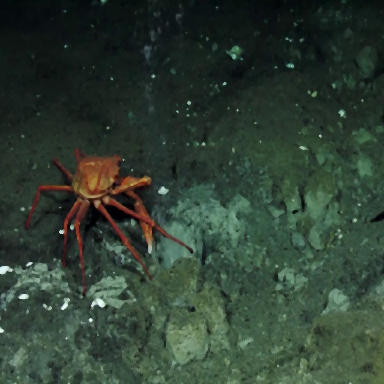}}
    \hfill
    \subfloat{\includegraphics[width=0.24\linewidth, height=0.24\linewidth]{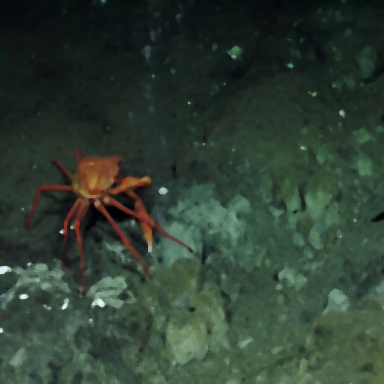}}
    \hfill
        \subfloat{\includegraphics[width=0.24\linewidth, height=0.24\linewidth]{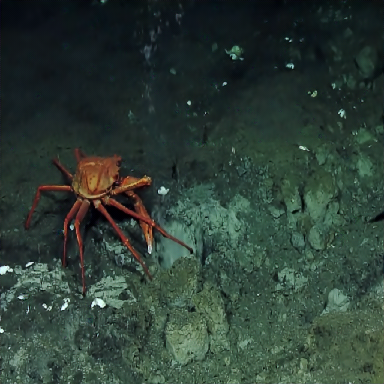}}
    \hfill

    \subfloat{\includegraphics[width=0.24\linewidth, height=0.24\linewidth]{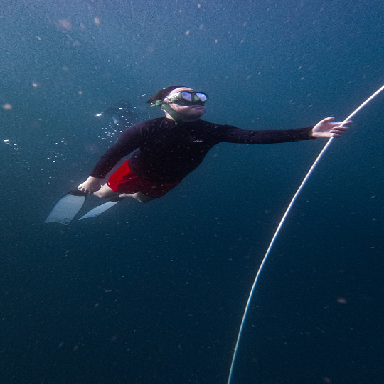}\label{subfig:oui}}
    \hfill
    \subfloat{\includegraphics[width=0.24\linewidth, height=0.24\linewidth]{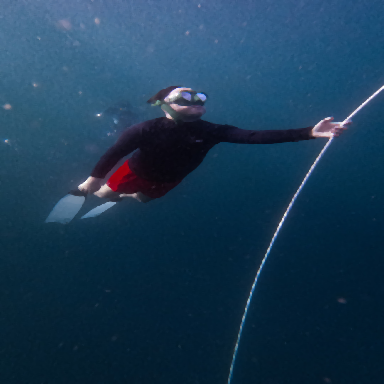}}
    \hfill
    \subfloat{\includegraphics[width=0.24\linewidth, height=0.24\linewidth]{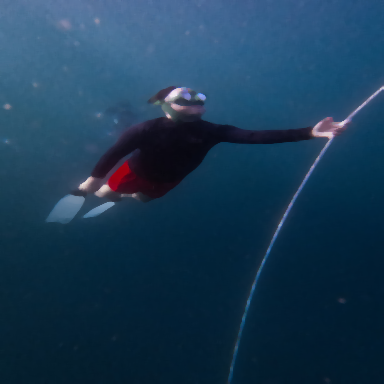}}
    \hfill
    \subfloat{\includegraphics[width=0.24\linewidth, height=0.24\linewidth]{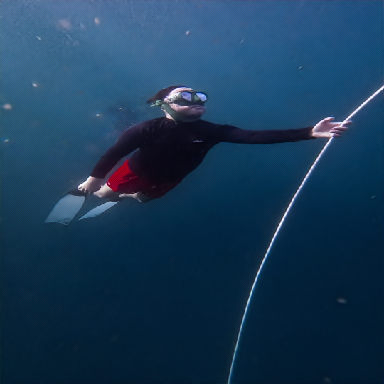}}
    \hfill

        \subfloat{\includegraphics[width=0.24\linewidth, height=0.24\linewidth]{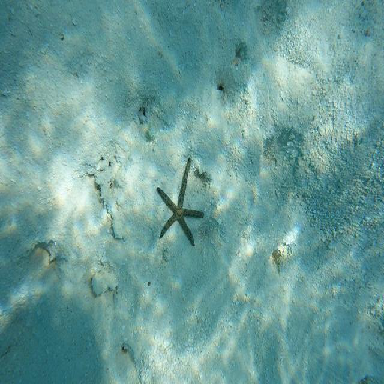}\label{subfig:o..i}}
    \hfill
    \subfloat{\includegraphics[width=0.24\linewidth, height=0.24\linewidth]{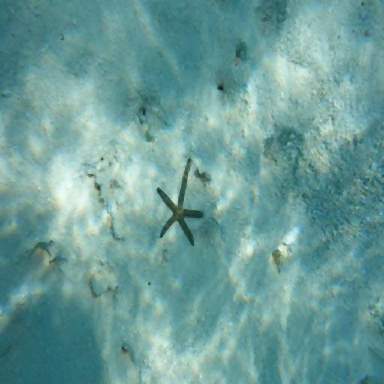}}
    \hfill
    \subfloat{\includegraphics[width=0.24\linewidth, height=0.24\linewidth]{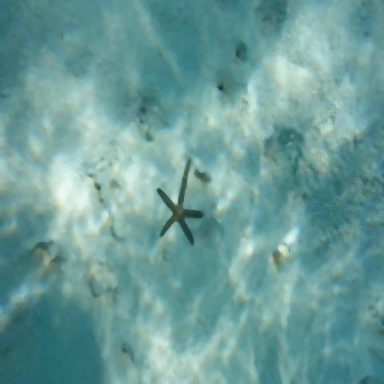}}
    \hfill
    \subfloat{\includegraphics[width=0.24\linewidth, height=0.24\linewidth]{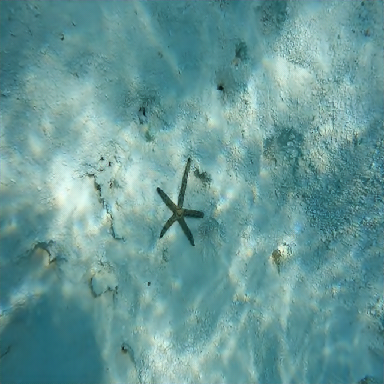}}
    \hfill

    \subfloat{\includegraphics[width=0.24\linewidth, height=0.24\linewidth]{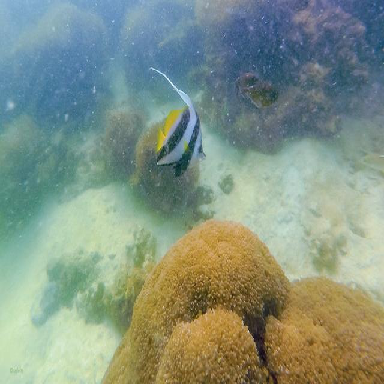}\label{subfig:oooi}}
    \hfill
    \subfloat{\includegraphics[width=0.24\linewidth, height=0.24\linewidth]{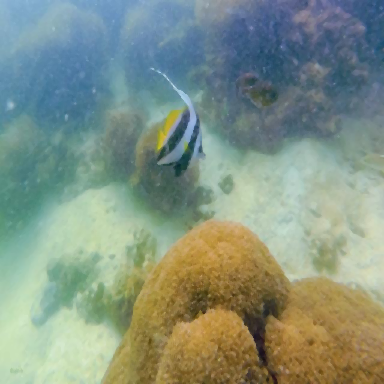}}
    \hfill
    \subfloat{\includegraphics[width=0.24\linewidth, height=0.24\linewidth]{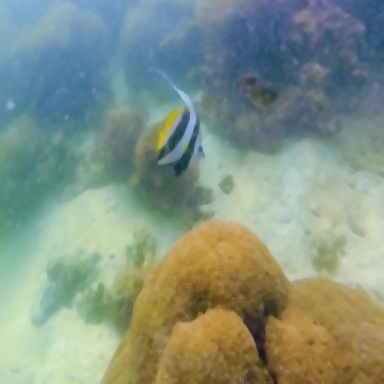}}
    \hfill
    \subfloat{\includegraphics[width=0.24\linewidth, height=0.24\linewidth]{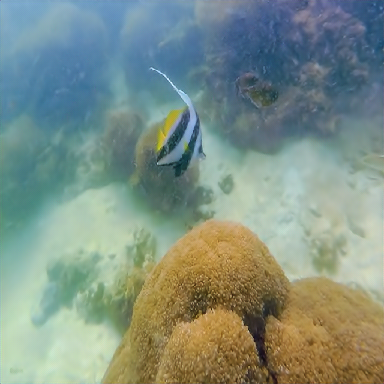}}
    \hfill

    \subfloat[Input]{\includegraphics[width=0.24\linewidth, height=0.24\linewidth]{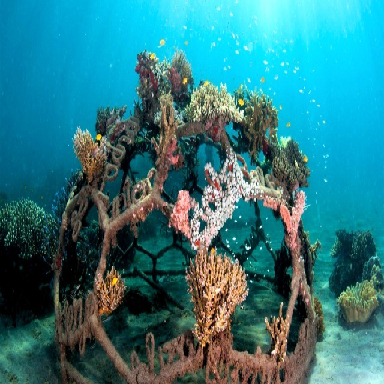}\label{subfig:oiW}}
    \hfill
    \subfloat[Median $3\times3$]{\includegraphics[width=0.24\linewidth, height=0.24\linewidth]{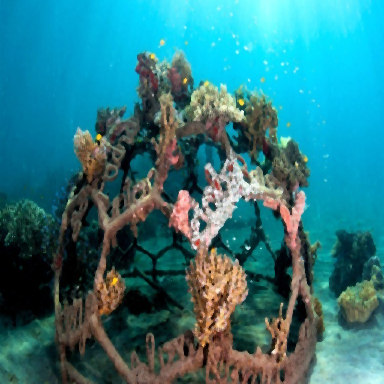}}
    \hfill
    \subfloat[Median $5\times5$]{\includegraphics[width=0.24\linewidth, height=0.24\linewidth]{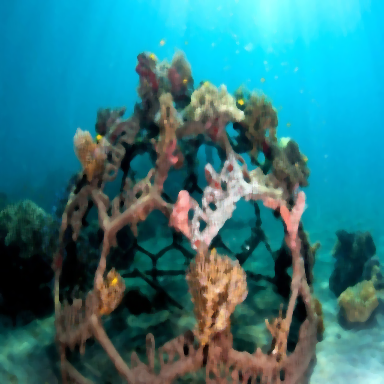}}
    \hfill
    \subfloat[Proposed]{\includegraphics[width=0.24\linewidth, height=0.24\linewidth]{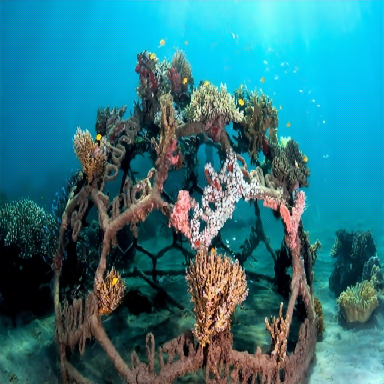}}
    \hfill

    \caption{Marine snow removal, additional results}
    \label{fig:resultsadditional2}
\end{figure*}
\begin{figure*}[t]
     \subfloat{\includegraphics[width=0.24\linewidth, height=0.24\linewidth]{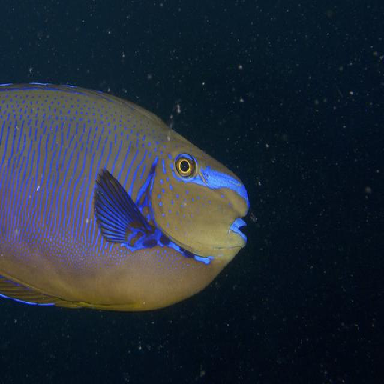}\label{subfig:oQi}}
    \hfill
    \subfloat{\includegraphics[width=0.24\linewidth, height=0.24\linewidth]{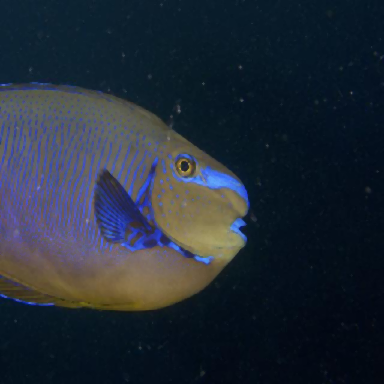}}
    \hfill
    \subfloat{\includegraphics[width=0.24\linewidth, height=0.24\linewidth]{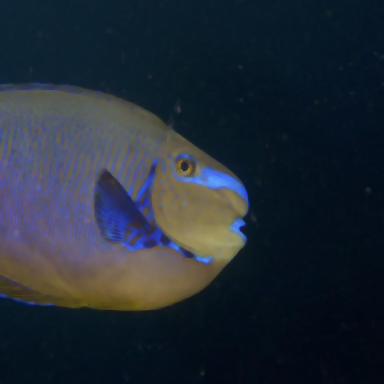}}
    \hfill
        \subfloat{\includegraphics[width=0.24\linewidth, height=0.24\linewidth]{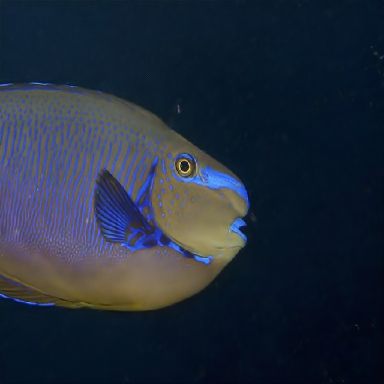}}
    \hfill

         \subfloat{\includegraphics[width=0.24\linewidth, height=0.24\linewidth]{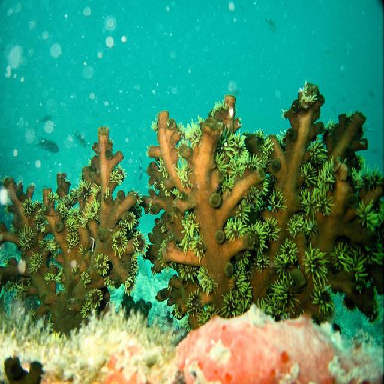}\label{subfig:qqetoi}}
    \hfill
    \subfloat{\includegraphics[width=0.24\linewidth, height=0.24\linewidth]{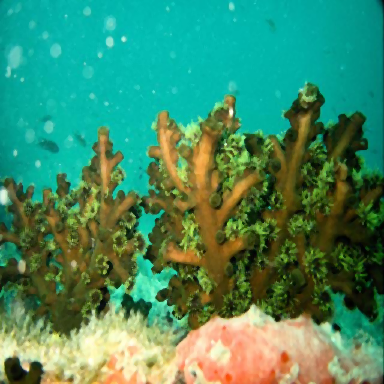}}
    \hfill
    \subfloat{\includegraphics[width=0.24\linewidth, height=0.24\linewidth]{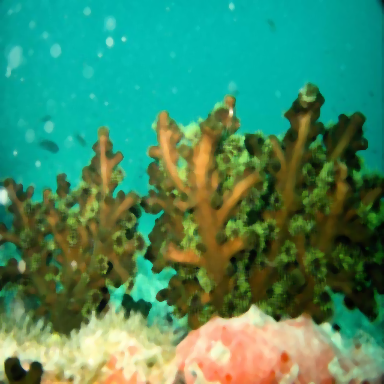}}
    \hfill
        \subfloat{\includegraphics[width=0.24\linewidth, height=0.24\linewidth]{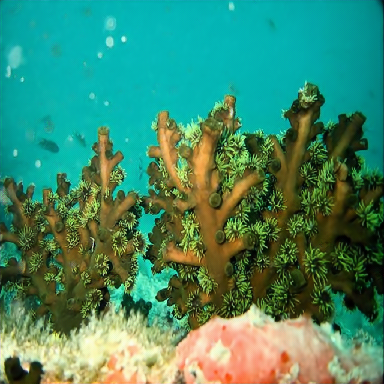}}
    \hfill

         \subfloat{\includegraphics[width=0.24\linewidth, height=0.24\linewidth]{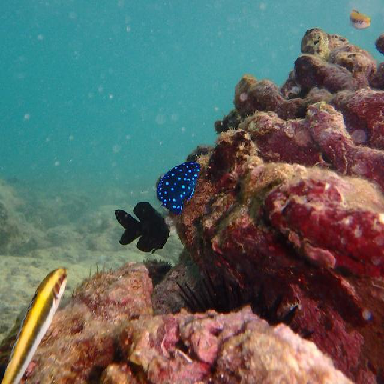}\label{subfig:otei}}
    \hfill
    \subfloat{\includegraphics[width=0.24\linewidth, height=0.24\linewidth]{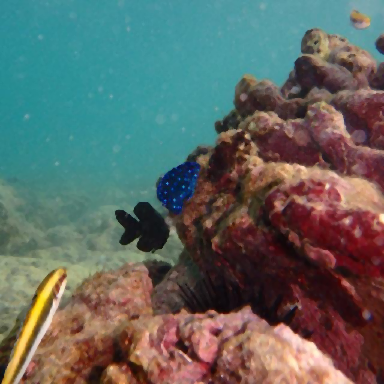}}
    \hfill
    \subfloat{\includegraphics[width=0.24\linewidth, height=0.24\linewidth]{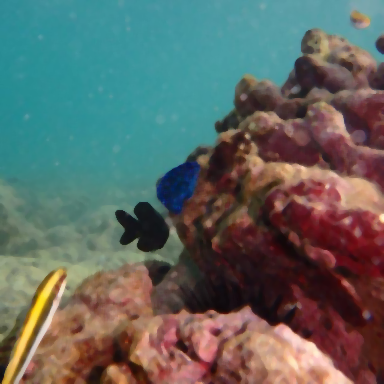}}
    \hfill
        \subfloat{\includegraphics[width=0.24\linewidth, height=0.24\linewidth]{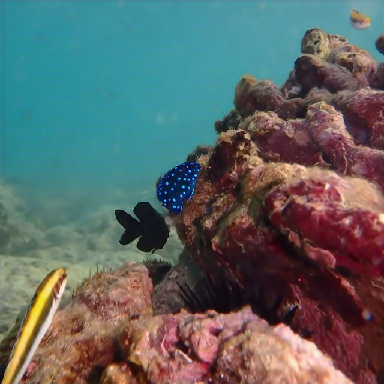}}
    \hfill

         \subfloat{\includegraphics[width=0.24\linewidth, height=0.24\linewidth]{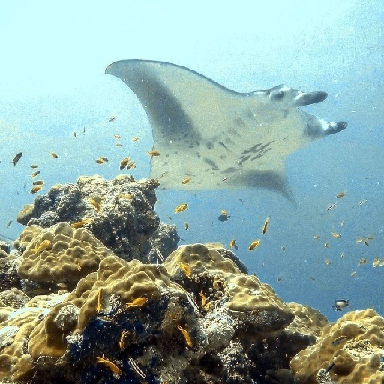}\label{subfig:hhoi}}
    \hfill
    \subfloat{\includegraphics[width=0.24\linewidth, height=0.24\linewidth]{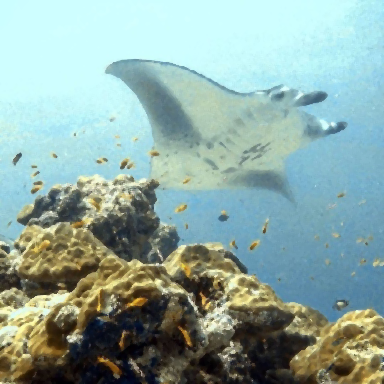}}
    \hfill
    \subfloat{\includegraphics[width=0.24\linewidth, height=0.24\linewidth]{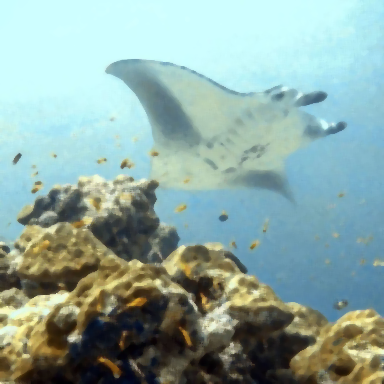}}
    \hfill
        \subfloat{\includegraphics[width=0.24\linewidth, height=0.24\linewidth]{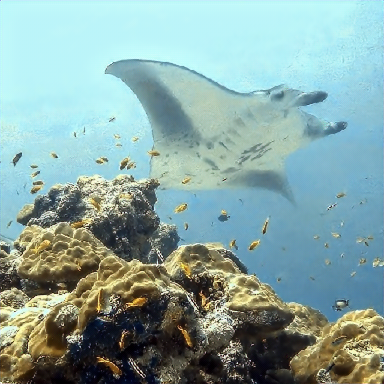}}
    \hfill

         \subfloat{\includegraphics[width=0.24\linewidth, height=0.24\linewidth]{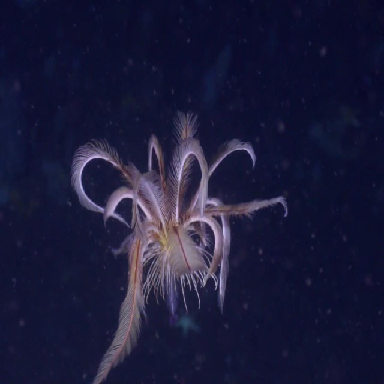}\label{subfig:obi}}
    \hfill
    \subfloat{\includegraphics[width=0.24\linewidth, height=0.24\linewidth]{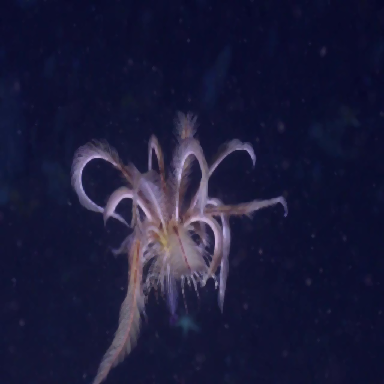}}
    \hfill
    \subfloat{\includegraphics[width=0.24\linewidth, height=0.24\linewidth]{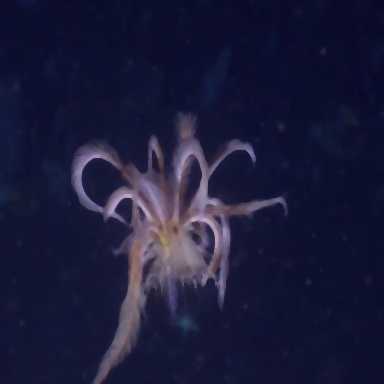}}
    \hfill
        \subfloat{\includegraphics[width=0.24\linewidth, height=0.24\linewidth]{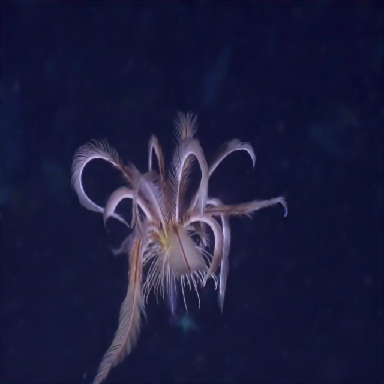}}
    \hfill

         \subfloat[Input]{\includegraphics[width=0.24\linewidth, height=0.24\linewidth]{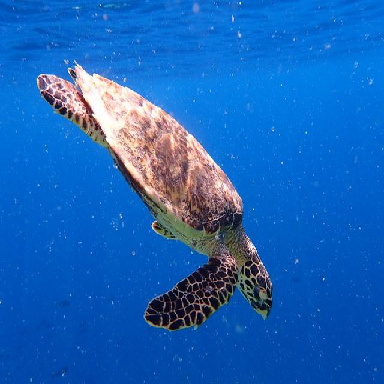}\label{subfig:odkdhi}}
    \hfill
    \subfloat[Median $3\times3$]{\includegraphics[width=0.24\linewidth, height=0.24\linewidth]{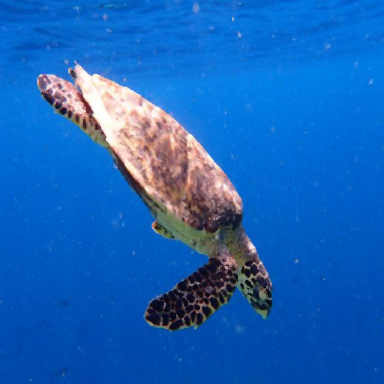}}
    \hfill
    \subfloat[Median $5\times5$]{\includegraphics[width=0.24\linewidth, height=0.24\linewidth]{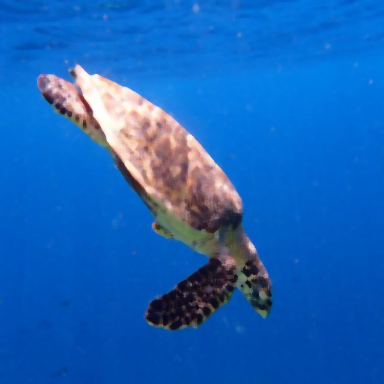}}
    \hfill
        \subfloat[Proposed]{\includegraphics[width=0.24\linewidth, height=0.24\linewidth]{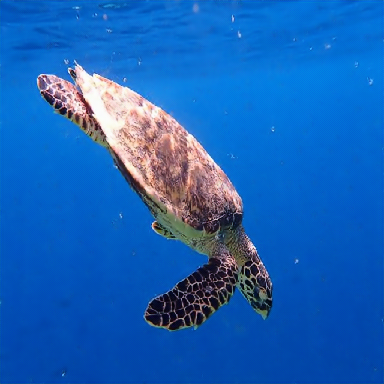}}
    \hfill

    \caption{Marine snow removal, additional results}
    \label{fig:resultsadditional4}
\end{figure*}